\documentclass[journal, a4paper]{IEEEtran}
\usepackage{amssymb}
\usepackage{amsfonts}
\usepackage{eucal}
\usepackage{bbm}
\usepackage{amsfonts}
\usepackage[dvips]{graphicx}
\usepackage{graphicx}
\usepackage{amsmath}
\usepackage{fancyhdr}
\usepackage{stfloats}
\usepackage{multirow}
\usepackage{algorithm}
\usepackage{algorithmic}
\usepackage{cite}
\usepackage[bookmarks=false]{}
\usepackage{amsthm}
\usepackage{algorithm}
\usepackage{algorithmic}
\usepackage{mathbbol}
\usepackage{amsfonts}
\usepackage{graphicx}
\usepackage{amsmath}
\usepackage{amssymb}
\usepackage{latexsym}
\usepackage{graphicx}
\usepackage{cite}
\usepackage{url}
\usepackage{stfloats}
\usepackage{mathrsfs}
\usepackage{setspace}
\usepackage{booktabs}
\usepackage {latexsym}
\usepackage {graphicx}
\usepackage {cite}
\usepackage {url}
\usepackage {stfloats}
\usepackage {mathrsfs}
\theoremstyle{plain}

\usepackage{cite}
\usepackage{setspace}
\ifCLASSINFOpdf
\usepackage[pdftex]{graphicx}
\else
\fi
\usepackage[tight,footnotesize]{subfigure}
\begin{document}
\clearpage
\title{\huge Optimization of User Selection and Bandwidth Allocation for Federated Learning in VLC/RF Systems}
%
\author{\IEEEauthorblockN{Chuanhong Liu\IEEEauthorrefmark{1}, Caili Guo\IEEEauthorrefmark{2}, Yang Yang\IEEEauthorrefmark{1}, Mingzhe Chen{\IEEEauthorrefmark{3}}, H. Vincent Poor\IEEEauthorrefmark{3}, and Shuguang Cui\IEEEauthorrefmark{4}\IEEEauthorrefmark{5}
\vspace{0.3cm}
\\\IEEEauthorrefmark{1}\small Beijing Key Laboratory of Network System Architecture and Convergence, School of
Information and Communication Engineering, Beijing University of Posts and Telecommunications, Beijing 100876, China
\\Emails: \{2016\_liuchuanhong, yangyang01\}@bupt.edu.cn.
\\\IEEEauthorrefmark{2}\small Beijing Laboratory of Advanced Information Networks, School of
Information and Communication Engineering, Beijing University of Posts and Telecommunications, Beijing 100876, China, Email: guocaili@bupt.edu.cn.
\\\IEEEauthorrefmark{3}\small Department of Electrical Engineering, Princeton University, Princeton, NJ, 08544, USA,
\\Emails: \{mingzhec, poor\}@princeton.edu.
\\\IEEEauthorrefmark{4}\small The Future Network of Intelligence Institute, The Chinese University of Hong Kong, Shenzhen, 518172, China.
\\\IEEEauthorrefmark{5}\small Shenzhen Research Institute of Big Data, The Chinese University of Hong Kong, Shenzhen, 518172, China,
\\Email: shuguangcui@cuhk.edu.cn.
}

\thanks{ This work was supported in part by National Natural Science Foundation of China (61871047), National Natural Science Foundation of China (61901047), Beijing Natural Science Foundation (4204106) and China Postdoctoral Science Foundation (2018M641278), and in part by the U.S. National Science Foundation under Grant CCF-1908308, and in part by the Shenzhen Outstanding Talents Training Fund, and by Guangdong Research Project No. 2017ZT07X152.
}
}
%
%
%
\maketitle
\thispagestyle{empty}
%
\begin{abstract}
Limited radio frequency (RF) resources restrict the number of users that can participate in federated learning (FL) thus affecting FL convergence speed and performance. In this paper, we first introduce visible light communication (VLC) as a supplement to RF in FL and build a hybrid VLC/RF communication system, in which each indoor user can use both VLC and RF to transmit its FL model parameters. Then, the problem of user selection and bandwidth allocation is studied for FL implemented over a hybrid VLC/RF system aiming to optimize the FL performance. The problem is first separated into two subproblems. The first subproblem is a user selection problem with a given bandwidth allocation, which is solved by a traversal algorithm. The second subproblem is a bandwidth allocation problem with a given user selection, which is solved by a numerical method. The final user selection and bandwidth allocation are obtained by iteratively solving these two subproblems. Simulation results show that the proposed FL algorithm that efficiently uses VLC and RF for FL model transmission can improve the prediction accuracy by up to 10\% compared with a conventional FL system using only RF.
\end{abstract}


\vspace{-0.2cm}
\section{Introduction}
\label{sec:intro}
\IEEEPARstart{F}{ederated} learning (FL), which allows edge devices to cooperatively train a shared machine learning model without the direct transmission of private data, is an emerging distributed machine learning technique \cite{fl1,fl2}. During the FL training process, the model parameters need to be transmitted iteratively over wireless links. Due to the dynamic wireless channels and imperfect wireless transmission, the performance of FL will be significantly affected by wireless communication and the number of users that can participate in FL is limited. Therefore, it is necessary to optimize wireless network performance for improving FL performance and convergence speed.

A number of prior studies have considered the optimization of FL over wireless networks. These studies include the optimization of energy efficiency and communication cost \cite{cost1,cost2,cost3}, framework design and user selection \cite{framework1,framework2,user1,user2,scheduling,convergence}.
 Among these studies, user selection is one of the most challenging problems. This is because only a subset of users can participate in FL training due to the limited bandwidth resources, thus significantly affecting the FL performance. The authors in \cite{user2} provided a general introduction to FL and demonstrated that user selection is one of the challenges in FL due to the limited wireless bandwidth and users' computation ability.
The work in \cite{framework1} showed that the number of users that participate in FL will significantly affect the performance of the trained model. A new client selection protocol referred to as federated learning with client selection (FedCS) was presented in \cite{user1}. However, the protocol in \cite{user1} cannot increase the number of selected users since the wireless radio frequency (RF) bandwidth is limited. The authors in \cite{scheduling} analyzed the effect of three practical scheduling policies on the performance of federated learning. We note that all of the existing works investigated the optimization of FL performance in RF only systems, which limits the number of devices that can participate in FL and thus affecting the FL performance. Visible light communication (VLC) can provide large, license-free bandwidth, hence, it can be a complement to RF in FL. Moreover, there is no interference between the RF and VLC systems, a key benefit of introducing VLC to future heterogeneous networks. Based on this observation, this work introduces VLC to enhance the capability of a conventional RF network to support FL, and investigates the FL performance optimization in the introduced hybrid VLC/RF system.

The main contribution of this paper is a joint user selection and bandwidth allocation algorithm that minimizes the training loss of FL implemented over the proposed hybrid VLC/RF system. To our best knowledge, \emph{this is the first work that introduces the use of VLC technique for FL performance optimization.} The contributions are summarized as follows:
\vspace{-0.3cm}
\begin{itemize}
\item[$\bullet$] We introduce VLC into conventional RF systems to improve FL performance. In the hybrid VLC/RF system, the bandwidth of RF and VLC must be appropriately allocated, which enables more users to participate in FL training. A joint user selection and bandwidth allocation problem is formulated, whose goal is to minimize the FL training loss.
\item[$\bullet$] To solve this problem, we first separate it into two sub-problems. The first subproblem is a user selection problem with a given bandwidth allocation, which is solved by a traversal algorithm. Based on the obtained subset of selected users, the second subproblem is to find the optimal bandwidth allocation, which is solved by a numerical method. The two subproblems are then updated iteratively until a convergent solution is obtained.
\end{itemize}

Simulation results verify that the proposed algorithm in a hybrid system can obtain 20\% and 10\% gains in terms of the number of selected users and the model accuracy, respectively, when compared with a conventional FL system using only RF.

The remainder of this paper is organized as follows. In Section II, we introduce the hybrid VLC/RF system. Section III introduces the system model used in this work. The joint user selection and bandwidth allocation algorithm is described in Section IV. Simulation and numerical results are presented and discussed in Section V. Finally, Section VI draws some important conclusions.


\vspace{-0.3cm}
\section{Design of Hybrid VLC/RF Systems}
\vspace{-0.1cm}
\label{sec:system}
In this section, we first review a traditional FL model based on an RF system and then summarize some challenges for training FL. To overcome these challenges, we then design a hybrid VLC/RF system for FL.
\vspace{-0.5cm}
\subsection{FL based on RF system}
\vspace{-0.1cm}
In this model, each user $n$ stores a local dataset ${{{\cal D}}_n}$ with $D_n$ being the number of training data samples. Hence, the total number of training data samples of all users is $D = \sum\nolimits_{n = 1}^N {{D_n}}$.
We assume that training data of user $n$ can be expressed by  $\{ {{\bf{x}}_n},{{\bf{y}}_n}\} $ with ${{\bf{x}}_n} = [{{\bf{x}}_{n1}},...{{\bf{x}}_{n{D_n}}}]$ and ${{\bf{y}}_n} = [{{\bf{y}}_{n1}},...{{\bf{y}}_{n{D_n}}}]$, where each element ${{\bf{x}}_{ni}}$ is an input vector of the FL algorithm and ${{\bf{y}}_{ni}}$ is the output of  ${{\bf{x}}_{ni}}$.

 For user $n$, the FL training purpose is to find the model parameter ${\boldsymbol{\omega }}$ that minimizes the loss function:
\begin{eqnarray}
{J_n}({\boldsymbol{\omega }}): = \frac{1}{{{D_n}}}\sum\nolimits_{i \in {{{\cal D}}_n}} {{f_{ni}}({\boldsymbol{\omega }})},
\end{eqnarray}
where ${f_{ni}}({\boldsymbol{\omega }})$ is a loss function that captures the performance of the FL algorithm. For example, for a linear regression FL, the loss function is ${f_{ni}}({\boldsymbol{\omega }}) = \frac{1}{2}{({{\bf{x}}_{ni}}^T{\boldsymbol{\omega }} - {{\bf{y}}_{ni}})^2}$ \cite{framework2}.

Then, the goal is to minimize the following global loss function:
\begin{eqnarray}
\label{e2}
\mathop {\min }\limits_{{\boldsymbol{\omega }}} J({\boldsymbol{\omega }}): = \sum\limits_{n = 1}^N {\frac{{{D_n}}}{D}} {J_n}({\boldsymbol{\omega }}).
\end{eqnarray}

To solve (\ref{e2}), the BS will transmit the global FL model parameters to its users and users will use the received global FL model parameters to train their local FL models. Then, the users will transmit their trained local FL model parameters to the BS to update the global FL model. For strongly convex objective $J({\boldsymbol{\omega }})$, the general upper bound on global iterations is\cite{complexity}
\begin{eqnarray}
K(\varepsilon ,\theta ) = \frac{{o(\log ({\raise0.7ex\hbox{$1$} \!\mathord{\left/
 {\vphantom {1 \varepsilon }}\right.\kern-\nulldelimiterspace}
\!\lower0.7ex\hbox{$\varepsilon $}}))}}{{1 - \theta }},
\end{eqnarray}
where $\varepsilon$ is the accuracy of global model and $\theta$ is the accuracy of local model. On the other hand, each global iteration consists of both computational and communication time. We consider a fixed global accuracy $\varepsilon$. Besides,
${o(\log ({\raise0.7ex\hbox{$1$} \!\mathord{\left/
 {\vphantom {1 \varepsilon }}\right.\kern-\nulldelimiterspace}
\!\lower0.7ex\hbox{$\varepsilon $}}))}$ is normalized to 1 so that $K(\theta ) = \frac{1}{{1 - \theta }}$ for ease of presentation.

Due to the limited wireless bandwidth, only a subset of users can be selected for FL training, which can seriously degrade the training accuracy. To enable more users to join the FL training process, we design a hybrid VLC/RF system.

\vspace{-0.4cm}
\subsection{FL based on hybrid VLC/RF system}
\begin{figure}[t]
\setlength{\abovecaptionskip}{-0.2cm}
\centering
\includegraphics[width=0.3\textwidth]{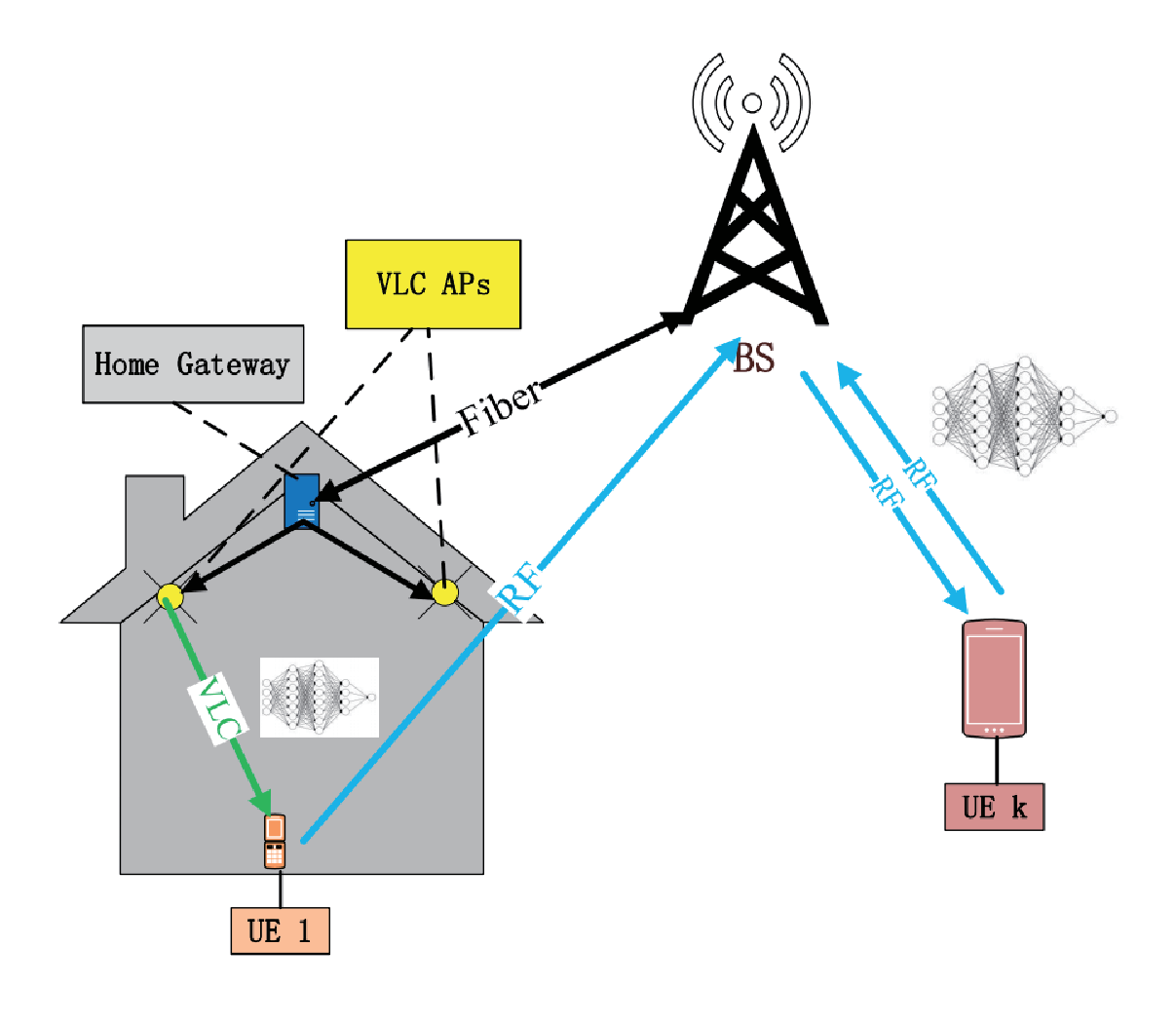}
\caption{Illustration of FL based on hybrid VLC/RF system.}
\label{fig:system}
\vspace{-0.2cm}
\end{figure}

\vspace{-0.1cm}
Consider a cellular network that consists of one BS, home gateways, and users cooperatively performing an FL algorithm for data analysis and inference. Denote the total users by a set ${{\cal N}}$ of $N$ users. Denote the indoor users by a set ${{\cal N}_1}$ of $N_1$ users and the outdoor users by a set ${{\cal N}_2}$ of $N_2$ users. In this work, we do not consider the mobility of users for simplicity. The system architecture is shown in Fig. \ref{fig:system}. In this model, the BS will send the global FL model parameters to outdoor users by RF. Meanwhile, the BS transmits the global model parameters to the home gateways which are connected to the indoor VLC access points (APs). Then, the VLC APs transmit the global FL model parameters to indoor users through the visible light signal. Assuming that the BS and home gateways are connected by fiber on which the bit error can be negligible.

\vspace{-0.cm}

In indoor scenarios, each VLC AP consists of an LED lamp. Each user is served by the AP that provides the strongest signal. In addition, we assume that all indoor users can be covered by visible lights. It is assumed that there is a central unit (CU) which controls both VLC and RF systems. Note that there is no interference between the RF and VLC systems, which is a key benefit of introducing VLC to the future heterogeneous networks.

\vspace{-0.1cm}
\section{System model and problem formulation}
\vspace{-0.1cm}
In this section, we elaborate the time and energy consumption
for training FL. In particular, the computational model is
first given. Then the communication models of both RF and
VLC are detailed. Finally, based on the established model, we
formulate a user selection and bandwidth allocation problem
in the proposed hybrid VLC/RF system.
\vspace{-0.4cm}
\subsection{Computational model}
\vspace{-0.1cm}
Let ${c_n}$ be the number of CPU cycles for user $n$ to process one sample of data. Since the data size of each training data sample is equal, the number of CPU cycles required for user $n$ to execute one local iteration is ${c_n}{D_n}$. Denote the CPU-cycle frequency of user $n$ by $f_n$. Then the CPU computational energy consumption of user $n$ in one global iteration can be expressed as follows:
\begin{eqnarray}
E_n^{cmp} = \sum\nolimits_{i = 1}^{{}{c_n}{D_n}} {\frac{{{\alpha _n}}}{2}} {f_n}^2 = \frac{{\nu {\alpha _n}{}{c_n}{D_n}}}{2}{f_n}^2\log (1/\theta ),
\end{eqnarray}
where $n = 1,2,...,N$, and ${\frac{{{\alpha _n}}}{2}}$ is the effective capacitance coefficient of the computing chipset of user $n$, and $\nu$ is a positive constant that depends on the data size of training data sample and the number of conditions in the local problem \cite{framework2}.

Furthermore, the computation time per local iteration of user $n$ can be denoted as $\frac{{{c_n}{D_n}}}{{{f_n}}},n = 1,2,...,N$. The computation time, however, depends on the number of local iterations, which is upper bounded by $o(\log (1/\theta ))$. Therefore, the required
computation time of user $n$ for data processing is
\begin{eqnarray}
t_n^{cmp} = \frac{{\nu {c_n}{D_n}\log (1/\theta )}}{{{f_n}}}.
\end{eqnarray}

\vspace{-0.5cm}
\subsection{RF Transmission Model}
\vspace{-0.1cm}
We use the orthogonal frequency division multiple access (OFDMA) technique for both uplink and downlink RF transmissions. The uplink rate of user $n$ is given by
\begin{eqnarray}
r_n^U = \sum\limits_{i = 1}^{{R^U}} {r_{n,i}^U{B^U}{{\log }_2}(1 + \frac{{{P_n}{h_n}}}{{\sum\limits_{{i^{'}} \in {{\cal U}}_n^{'}} {{P_{{i^{'}}}}} {h_{{i^{'}}}} + {B^U}N_0^\emph{RF}}})},
\end{eqnarray}
where ${\bf{r}}_n^U = [r_{n,1}^U,...,r_{n,R^U}^U]$ is a resource block (RB) allocation vector and ${{R^U}}$ is the total number of RBs allocated to RF uplink; $r_{n,i}^U \in \{ 0,1\}$ and $\sum\limits_{i = 1}^{{R^U}} {r_{n,i}^U = 1}$; ${r_{n,i}^U = 1}$ implies that RB $i$ is allocated to user $n$; otherwise, we have ${r_{n,i}^U = 0}$; ${{{\cal U}}_n^{'}}$ represents the set of users that are located at the other service
areas and transmit data over RB $i$; ${{B^U}}$ is the bandwidth of each RB and $P_n$ is the transmit power of user $n$; $h_n$ is the channel gain between user $n$ and the BS; $N_0^\emph{RF}$ is the noise power spectral density; ${\sum\limits_{{i^{'}} \in {{\cal U}}_n^{'}} {{P_{{i^{'}}}}} {h_{{i^{'}}}}}$ is the interference caused by the users that are located in other service areas and use the same RB.

On the other hand, the downlink data rate achieved by the BS to each user $n$ is given by
\begin{eqnarray}
r_n^D = \sum\limits_{i = 1}^{{R^D}} {r_{n,i}^D{B^D}{{\log }_2}(1 + \frac{{{P_B}{h_n}}}{{\sum\limits_{j \in {{{\cal B}}^{'}}} {{P_B}} {h_{nj}} + {B^D}N_0^\emph{RF}}})},
\end{eqnarray}
where ${{B^D}}$ is the bandwidth of each RB that the BS used to transmit the global FL model to each user $n$; ${\bf{r}}_n^D = [r_{n,1}^D,...,r_{n,{R^{D}}}^D]$ is a RB allocation vector with $R^D$ being the total number of RBs allocated to RF downlink, $r_{n,i}^D \in \{ 0,1\}$ and $\sum\limits_{i = 1}^{{R^D}} {r_{n,i}^D = 1}$; ${r_{n,i}^D = 1}$ indicates that RB $i$ is allocated to user $n$; otherwise, we have ${r_{n,i}^D = 0}$; ${{P_B}}$ is the transmit power of the BS; ${{{{\mathcal B}}^{'}}}$ is the set of other BSs that cause interference to the BS that performs the FL algorithm; ${{h_{nj}}}$ is the channel gain between user $n$ and BS $j$. Let $B_R$ be the total RF bandwidth, and we have ${R^U} \times {B^U} + {R^D} \times {B^D} \le {B_R}$. For simplicity, we assume $B^U = B^D$ which means the bandwidth of uplink resource block is equal to that of downlink RB.

Denote the data size of the FL model that each user needs to upload by $s$. To upload the local FL model within transmission time $t_n^U$, we have $t_n^Ur_n^U \ge s$. Meanwhile, the required energy of user $n$ is $E_n^{com} = t_n^U{P_n}$. Similarly, to download the global FL model within transmission time $t_n^D$, we have $t_n^Dr_n^D \ge s$.

\addtolength{\topmargin}{0.7cm}

\vspace{-0.4cm}
\subsection{VLC Transmission Model}
\vspace{-0.1cm}
According to \cite{vlc1} and \cite{vlc2}, the optical channel gain of a line-of-sight (LoS) channel can be expressed as
\begin{eqnarray}
u = \left\{ {\begin{array}{*{20}{l}}
{\frac{{(m + 1){A_p}}}{{2\pi {d^2}}}{T_s}(\theta )g(\theta ){{\cos }^m}(\varphi )\cos (\theta ),0 < \theta  \le {\Theta _F}},\\
{0,\theta  > {\Theta _F},}
\end{array}} \right.
\end{eqnarray}
where $m =  - \frac{1}{{{{\log }_2}(\cos ({\theta _{1/2}}))}}$ is the Lambertian index which is a function of the half-intensity radiation angle ${{\theta _{1/2}}}$; ${{A_p}}$ is the receiver's physical area of the photo-diode; $d$ is the distance from the VLC AP to the optical receiver; $\varphi$ is the angle of irradiation and $\theta$ is the angle of incidence; ${{\Theta _F}}$ is the half angle of the receiver's file of view (FoV); ${{T_s}(\theta )}$ is the gain of the optical filter; and the concentrator gain ${g(\theta)}$ can be written as
\begin{eqnarray}
g(\theta ) = \left\{ {\begin{array}{*{20}{l}}
{\frac{{{n_0^2}}}{{{{\sin }^2}{\Theta _F}}}, 0 < \theta  \le {\Theta _F}},\\
{0, \theta  > {\Theta _F},}
\end{array}} \right.
\end{eqnarray}
where $n_0$ is the refractive index. For a given user $n$ connected to a VLC AP $k$, the signal-to-interference-plus-noise ratio (SINR) can be written as
\begin{eqnarray}
{s_{nk}} = \frac{{{{(\gamma {u_{nk}}{P_v})}^2}}}{{N_0^\emph{VLC}B + \sum\nolimits_{l \ne k} {{{(\gamma {u_{nl}}{P_v})}^2}} }},
\end{eqnarray}
where $\gamma$ is the optical to electric conversion efficiency; ${{P_v}}$ is the transmitted optical power of a VLC AP; ${N_0^\emph{VLC}}$ is the
noise power spectral density; ${{u_{nk}}}$ is the channel gain between user $n$ and the VLC AP $k$; ${{u_{nl}}}$ is the channel gain between user $n$ and the interfering VLC AP $l$; $B$ is the bandwidth of each VLC RB. Each user is served by a single VLC AP which has the largest SINR for the user. In the VLC system, optical OFDMA is employed. It is known
that the input signal of the LEDs is amplitude constrained.
Therefore, the classical Shannon capacity formula for complex
and average power constrained signal is not applicable in VLC.
Therefore, the lower bound of achievable data rate is used,
which can be expressed as \cite{vlc3}
\begin{eqnarray}
{r_n} = \sum\limits_{i = 1}^{{R^V}} {r_{n,i}^V\frac{B}{2}{{\log }_2}(1 + \frac{2}{{\pi e}}{s_n})},
\end{eqnarray}
where ${{s_n}}$ is the largest SINR which is evaluated as ${s_n} = \max \{ {s_{n1}},...,{s_{nK}}\}$, where $K$ is the total number of VLC APs; ${\bf{r}}_n^V = [r_{n,1}^V,...,r_{n,R^V}^V]$ is an RB allocation vector with $R^V$ being the total number of VLC RBs, $r_{n,i}^V \in \{ 0,1\}$ and $\sum\limits_{i = 1}^{{R^V}} {r_{n,i}^V = 1}$; ${r_{n,i}^V = 1}$ indicates that RB $i$ is allocated to user $n$; otherwise, we have ${r_{n,i}^V = 0}$. Similarly, we have ${R^V} \times B \le {B_V}$, where $B_V$ is the total bandwidth of VLC.

We assume that the data size of global FL model parameters which
are transmitted to users can also be denoted by $s$. Therefore,
the downlink communication time of indoor user $n$ in each
global iteration will be ${t_{dn}} = \frac{s}{{{r_n}}}$.

\vspace{-0.3cm}
\subsection{Problem Formulation}
\vspace{-0.1cm}
Our goal is to fully exploit the complementary
function of VLC systems and appropriately allocate
the precious bandwidth of both RF and VLC for enhancing
FL performance. To this end, we formulate an optimization
problem whose goal is to minimize the global loss function
under time, energy, and bandwidth
allocation constraints. The minimization problem is given by
\begin{align}
&{\mathop {\min }\limits_{B,{B^D},{B^U},{{\cal S}}} J({\boldsymbol{\omega }})}\\
\rm{s.\;t.}\;\;\;&{R^U} \times {B^U} + {R^D} \times {B^D} \le {B_R},\tag{\theequation a}\\
&{R^V} \times B \le {B_V},\tag{\theequation b}\\
&{t_{dn}} + t_n^U + t_n^{cmp} + {t_d} \le {T_{round,}}\forall n \in {{{\cal S}}_{1}},\tag{\theequation c}\\
&t_n^D + t_n^U + t_n^{cmp} \le {T_{round,}}\forall n \in {{{\cal S}}_2},\tag{\theequation d}\\
&{{{\cal S}}_{1}} \cup {{{\cal S}}_2} = {{\cal S}},\tag{\theequation e}\\
&E_n^{com} + E_n^{cmp} \le {\gamma _{nE}},\forall n \in {{\cal N}},\tag{\theequation f}\\
&{R^U} = \left| {{\cal S}} \right|,{R^D} = \left| {{{{\cal S}}_2}} \right|,{R^V} = \left| {{{{\cal S}}_1}} \right|,\tag{\theequation g}
\end{align}
where ${{\cal S}}$ denotes the set of selected users participating in FL, ${{{\cal S}}_{1}}$ denotes the set of selected indoor users, ${{{\cal S}}_{2}}$ denotes the set of selected outdoor users, and $\left| . \right|$ denotes the cardinality of a set. In addition, ${T_{round}}$ is the time threshold for each round and $t_d$ denotes the delay between the BS and the home gateway. In addition, ${\gamma _{nE}}$ is the energy constraint of user $n$. Constraint (12c) is the delay constraint of
each round for all selected indoor users while (12d) is the
delay constraint of each round. In addition, (12f) is the energy
consumption requirement of performing an FL algorithm.
\vspace{-0.1cm}
\section{The Proposed Algorithm}
\vspace{-0.1cm}
Next, we first analyze the optimization problem (12) so as to figure out how the user selection and bandwidth allocation affect the FL performance. Then, a joint user selection and
bandwidth allocation (USBA) algorithm is proposed to solve the optimization problem.

\vspace{-0.cm}
\noindent\textbf{$Lemma\ 1.$} The optimization problem (12) can be transformed into an optimization problem with the objective function of maximizing the total sample size of the selected users when the users' transmit power are fixed, which can be denoted as
\begin{eqnarray}
\begin{array}{l}
\mathop {\max }\limits_{B,{B^D},{B^U},{{\cal S}}} \sum\limits_{n = 1}^{{N_1}} {\sum\limits_{i = 1}^{{R^V}} {{D_n}{r_{n,i}^V} + } } \sum\limits_{n = {N_1} + 1}^N {\sum\limits_{i = 1}^{{R^D}} {{D_n}{r_{n,i}^D}} } \\
{\rm{s}}{\rm{.t. }}\;\;{\rm{(12a) - (12g)}}
\end{array}
\end{eqnarray}

\begin{proof}
Minimizing the global loss function is equivalent to minimize the gap between the global loss function $J({{\boldsymbol{\omega }}_t})$ at time $t$ and the optimal global loss function $J({{\boldsymbol{\omega }}^*})$. According to Theorem 1 in \cite{framework1}, the gap is caused by the packet error rate (PER) and number of selected users. Here, we do not consider the packet errors and hence, we have $q_i=0$. Using the same simplification method in \cite{framework1}, the optimization problem can be transformed to problem (13).
This ends the proof.
\end{proof}
\vspace{-0.2cm}

Since the problem in (13) is non-convex, we first divide (13) into two
subproblems, and then solve these two subproblems iteratively.
In particular, we first fix the bandwidth allocation and calculate
the optimal user selection. Then, the problem of bandwidth
allocation is formulated and solved with the obtained subset of the selected users. After certain iterations, the obtained user selection and
bandwidth allocation remain unchanged, and that means a
convergent solution of (13) is obtained.


\vspace{-0.4cm}
\subsection{Optimal User Selection}
\vspace{-0.1cm}
Given the bandwidth of each RB, (13) can be simplified as
\begin{align}
&\mathop {\max }\limits_{{{\cal S}}} \sum\limits_{n = 1}^{{N_1}} {\sum\limits_{i = 1}^{{R^V}} {{D_n}{r_{n,i}^V} + } } \sum\limits_{n = {N_1} + 1}^N {\sum\limits_{i = 1}^{{R^D}} {{D_n}{r_{n,i}^D}} }\\
\rm{s.\;t.}\;\;&{t_{dn}} + t_n^U + t_n^{cmp} + {t_d} \le {T_{round,}}\forall n \in {{{\cal S}}_{1}},\tag{\theequation a}\\
&t_n^D + t_n^U + t_n^{cmp} \le {T_{round,}}\forall n \in {{{\cal S}}_2},\tag{\theequation b}\\
&{{{\cal S}}_{1}} \cup {{{\cal S}}_2} = {{\cal S}},\tag{\theequation c}\\
&E_n^{com} + E_n^{cmp} \le {\gamma _{nE}},\forall n \in {{\cal N}},\tag{\theequation d}\\
&{R^U} = \left| {{\cal S}} \right|,{R^D} = \left| {{{{\cal S}}_2}} \right|,{R^V} = \left| {{{{\cal S}}_1}} \right|,\tag{\theequation e}
\end{align}
We can observe from (15) that if the bandwidth of each RB
is fixed, the subset of selected users is determined by the user's
computing power and channel condition. We denote the
algorithm that select users under fixed bandwidth allocation
by \textbf{GetS}(${B^U}$,${B^D}$,$B$), which is summarized in \textbf{Algorithm} \ref{algorithm1}.
\begin{algorithm}[t]
\small
\caption{\textbf{GetS}(${B^U}$,${B^D}$,$B$)}
\begin{algorithmic}[1]
\STATE \textbf{Input:} ${{{\cal N}}_{1}},{{{\cal N}}_2}$.
\FOR {$n \in {{{\cal N}}_{1}}$}
    \IF {${t_{dn}} + t_n^U + t_n^{cmp} + {t_d} \le {T_{round,}}$\textbf{and}$E_n^{com} + E_n^{cmp} \le {\gamma _{nE}}$}
        \STATE ${{\cal S}_1} \leftarrow n$
    \ENDIF
\ENDFOR
\FOR {$n \in {{{\cal N}}_{2}}$}
    \IF {$t_n^D + t_n^U + t_n^{cmp} \le {T_{round,}}$\textbf{and}$E_n^{com} + E_n^{cmp} \le {\gamma _{nE}}$}
        \STATE ${{\cal S}_2} \leftarrow n$
    \ENDIF
\ENDFOR
\RETURN ${{\cal S}_1}$, ${{\cal S}_2}$
\end{algorithmic}
\label{algorithm1}
\end{algorithm}

\vspace{-0.6cm}
\subsection{Optimal RB Bandwidth}
\vspace{-0.1cm}
With an obtained subset of users, we then need to find the optimal $B$, $B^U$, and $B^D$ that can further optimize the capability of the
hybrid VLC/RF system. Note that the larger the bandwidth of each
RB is, the smaller the delay can be, implying more users can
be potentially selected. Based on this observation, the optimal
RB bandwidth allocation is
\begin{align}
&\max {B^U}\\
\rm{s.\;t.}\;\;&{R^U} \times {B^U} + {R^D} \times {B^D} \le {B_R},\tag{\theequation a}\\
&{B^U} = {B^D},\tag{\theequation b}\\
&{R^U} = \left| {{\cal S}} \right|,{R^D} = \left| {{{{\cal S}}_2}} \right|,\tag{\theequation c}
\end{align}
and
\begin{align}
&\max B\\
s.t.\;\;&{R^V} \times B \le {B_V},\tag{\theequation a}\\
&{R^V} = \left| {{{{\cal S}}_1}} \right|.\tag{\theequation b}
\end{align}
\vspace{0.2cm}
\noindent\textbf{$Lemma\ 2.$} The maximum bandwidth of each RB can be obtained when ${R^U} \times {B^U} + {R^D} \times {B^D} = {B_R}$ and ${R^V} \times B = {B_V}$.
\vspace{-0.2cm}
\begin{proof}
We use the contradiction method to prove $Lemma\ 2$. First, we assume that maximum $B_0^U$, $B_0^D$, and ${B_0}$ exist when (16a) and (17a) are not equal. Hence, we have
\begin{eqnarray}
B_0^U{\rm{ = }}B_0^D < \frac{{{B_R}}}{{\left| {{\cal S}} \right| + \left| {{{{\cal S}}_2}} \right|}},
\end{eqnarray}
and
\begin{eqnarray}
{B_0} < \frac{{{B_V}}}{{\left| {{{{\cal S}}_1}} \right|}}.
\end{eqnarray}
However, when (16a) and (17a) are equal, $B_1^U$, $B_1^D$, and $B_1$ satisfy the following equations:
\begin{eqnarray}
B_1^U{\rm{ = }}B_1^D = \frac{{{B_R}}}{{\left| {{\cal S}} \right| + \left| {{{{\cal S}}_2}} \right|}},
\end{eqnarray}
and
\begin{eqnarray}
{B_1} = \frac{{{B_V}}}{{\left| {{{{\cal S}}_1}} \right|}}.
\end{eqnarray}
Obviously, $B_1^U{\rm{ = }}B_1^D > B_0^U{\rm{ = }}B_0^D$ and ${B_1} > {B_0}$, which contradicts the assumption.
This ends the proof.
\end{proof}

Therefore, we have
\begin{eqnarray}
{B^U} = {B^D} = \frac{{{B_R}}}{{{R^U} + {R^D}}} = \frac{{{B_R}}}{{\left| {{\cal S}} \right| + \left| {{{{\cal S}}_2}} \right|}},
\end{eqnarray}
and
\begin{eqnarray}
B = \frac{{{B_V}}}{{{R^V}}} = \frac{{{B_V}}}{{\left| {{{{\cal S}}_1}} \right|}}.
\end{eqnarray}


\vspace{-0.4cm}
\subsection{Iterative Solution}
Once we obtain the bandwidth of each RB, we can obtain the optimal subset of selected users. Accordingly,
we can obtain the optimal bandwidth allocation based on the
obtained selected users, which is denoted by \textbf{GetB}(${{\cal S}}$). Then,
the selected users can be updated again based on the bandwidth
allocation. The iteration ends when both the user selection
and bandwidth allocation remain fixed. Obviously, the algorithm can always reach convergence after a certain number of iterations. We summarize the
proposed USBA algorithm in \textbf{Algorithm} \ref{algorithm2}.
\vspace{-0.1cm}
\begin{algorithm}[t]
\small
\caption{USBA Algorithm.}
\begin{algorithmic}[1]
\STATE \textbf{Input:} ${{\cal S}} \leftarrow \{ \} ,{B_0},B_0^D,B_0^U.$
\STATE \textbf{function} \textbf{GetB}(${{\cal S}}$)
\STATE ${B^U} = {B^D} = \frac{{{B_R}}}{{\left| {{\cal S}} \right| + \left| {{{{\cal S}}_2}} \right|}}$
\STATE $B = \frac{{{B_V}}}{{{R^V}}} = \frac{{{B_V}}}{{\left| {{{{\cal S}}_1}} \right|}}$
\STATE \textbf{end function}
\STATE \textbf{function main()}
\STATE $n=1$
\STATE ${{{\cal S}}^0} \leftarrow$ \textbf{GetS}($B_0^U,B_0^D,{B_0}$)
\WHILE{1}
    \STATE $[B_n^U,B_n^D,{B_n}] \leftarrow $ \textbf{GetB}(${{{\cal S}}^{n - 1}}$)
    \STATE ${{{\cal S}}^n} \leftarrow $ \textbf{GetS}($[B_n^U,B_n^D,{B_n}]$)
    \IF {${{{\cal S}}^n} == {{{\cal S}}^{n - 1}}$ and $[B_n^U,B_n^D,{B_n}] =  = [B_{n - 1}^U,B_{n - 1}^D,{B_{n - 1}}]$}
        \STATE \textbf{break}
    \ENDIF
    \STATE $n = n + 1$
\ENDWHILE
\STATE \textbf{end function}
\end{algorithmic}
\label{algorithm2}
\end{algorithm}

\vspace{-0.1cm}
\section{Simulation Results and ANALYSIS}
Consider a circular network area having a radius $r = 50$ m with one BS at its center. There are $N = 50$ uniformly
distributed users, and 80\% of the users are in indoors and
20\% of them are in outdoors. The system specifications are
summarized in Table \uppercase\expandafter{\romannumeral1}. The dataset used to train the FL
algorithm is Boston housing dataset\footnote{http://lib.stat.cmu.edu/datasets/boston} that is randomly allocated
to users. The number of samples of each user is equal. The goal of the FL algorithm is to train a simple
Back Propagation (BP) neural network with only one hidden
layer composed of 10 neurons. For comparison, we also execute the FL in RF-only systems.
\begin{table}[t]
\small
\vspace{-0.1cm}
\centering
\caption{Simulation Parameters.}
\setlength{\abovecaptionskip}{-0.5cm}
\begin{tabular}{ccc}
\toprule
Parameter & Value \\
\hline
Transmitted optical power per VLC AP, ${{P_v}}$ & 9 W \\
Modulation bandwidth for LED lamp, $B$ & 40 MHz\\
The physical area of a PD, ${A_p}$ & 1 c{m$^2$}\\
Half-intensity radiation angle, ${\theta _{1/2}}$ & 60 deg.\\
Gain of optical filter, ${{T_s}(\theta )}$ & 1.0\\
Receiver FOV semiangle, ${{\Theta _F}}$ & 90 deg.\\
Refractive index, $n$ & 1.5\\
Optical to electric conversion efficiency, $\gamma $ & 0.53 A/W\\
Noise power spectral density, $N_0^{VLC}, N_0^{RF}$ & ${10^{ - 21}}$ A$^{2}$/Hz\\
RF total bandwidth, ${{B_R}}$ & 20 MHz\\
Transmit power of BS, ${P_B}$ & 1 W\\
The number of users, $N$ & 50\\
Delay requirement, ${T_{round}}$ & 2.5 s\\
Energy consumption requirement, ${\gamma _{nE}}$ & 2 J\\
Energy consumption coefficient, $\alpha $ & $2 \times {10^{ - 28}}$\\
Data size of FL model, $s$ & 1 Mb\\
\toprule
\end{tabular}
\vspace{-0.cm}
\end{table}

\begin{figure}[t]
\setlength{\abovecaptionskip}{-0.1cm}
\centering
\includegraphics[width=0.52\textwidth]{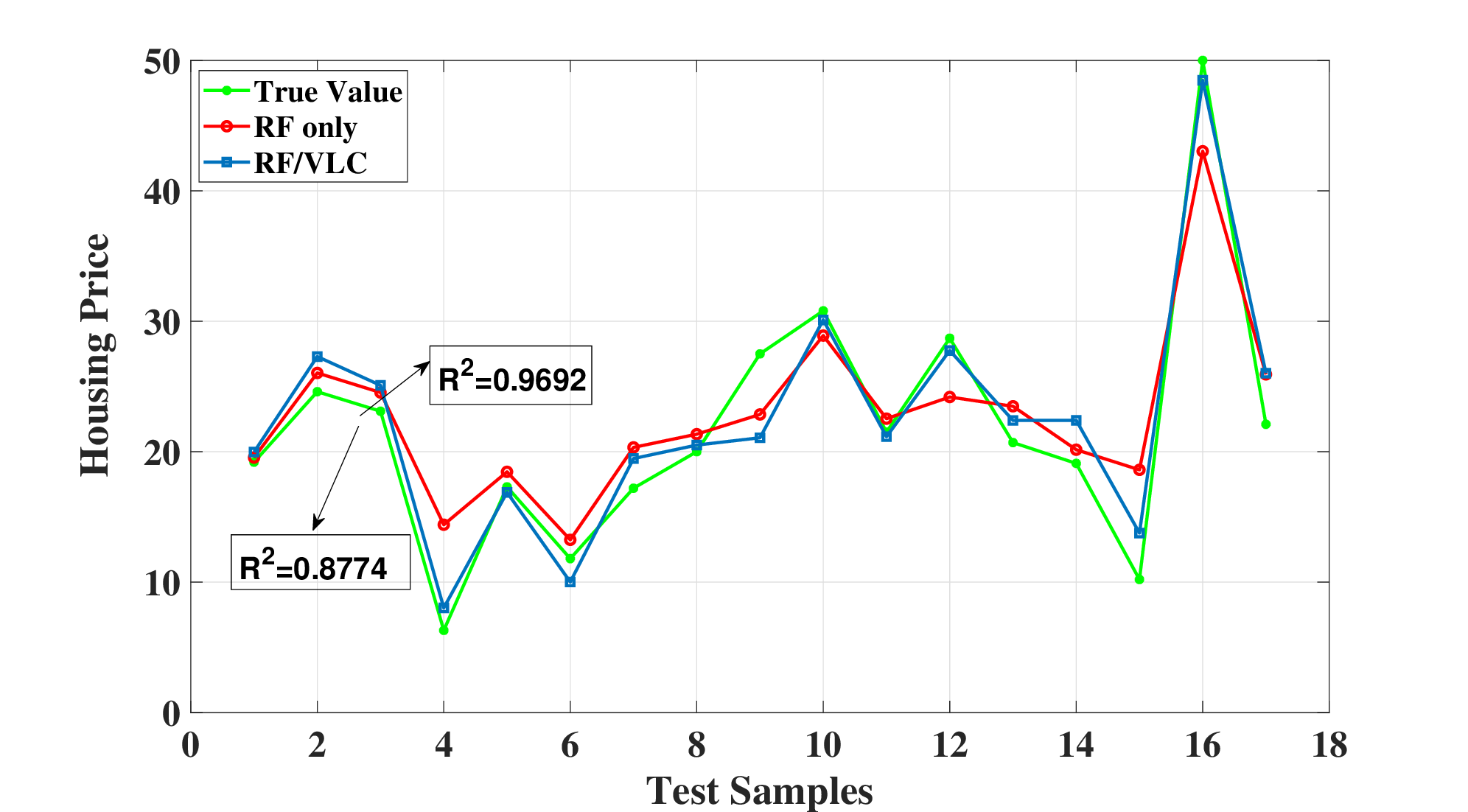}
\caption{An example of implementing FL for a BP neural network.}
\label{fig:housing}
\vspace{-0.4cm}
\end{figure}

As show in Fig. \ref{fig:housing}, the FL algorithm is used for predicting the housing price. In this figure, the green line is the true values of data samples. Before training, we randomly select 17 samples to
form a test set for testing. Here, we use the coefficient of
determination ($R^2$) to measure the quality of the model. The
higher value of $R^2$ is, the higher prediction accuracy is. From Fig. \ref{fig:housing}, we can observe that the proposed USBA algorithm can
achieve better performance than baseline. The $R^2$ has increased by 10\%
when compared with RF-only system. This is because the proposed USBA algorithm introduces visible light communication,
which can get higher communication rate and quality,
thereby increasing the number of selected users and further
improving the FL performance.

\begin{figure}[t]
\setlength{\abovecaptionskip}{-0.1cm}
\centering
\includegraphics[width=0.52\textwidth]{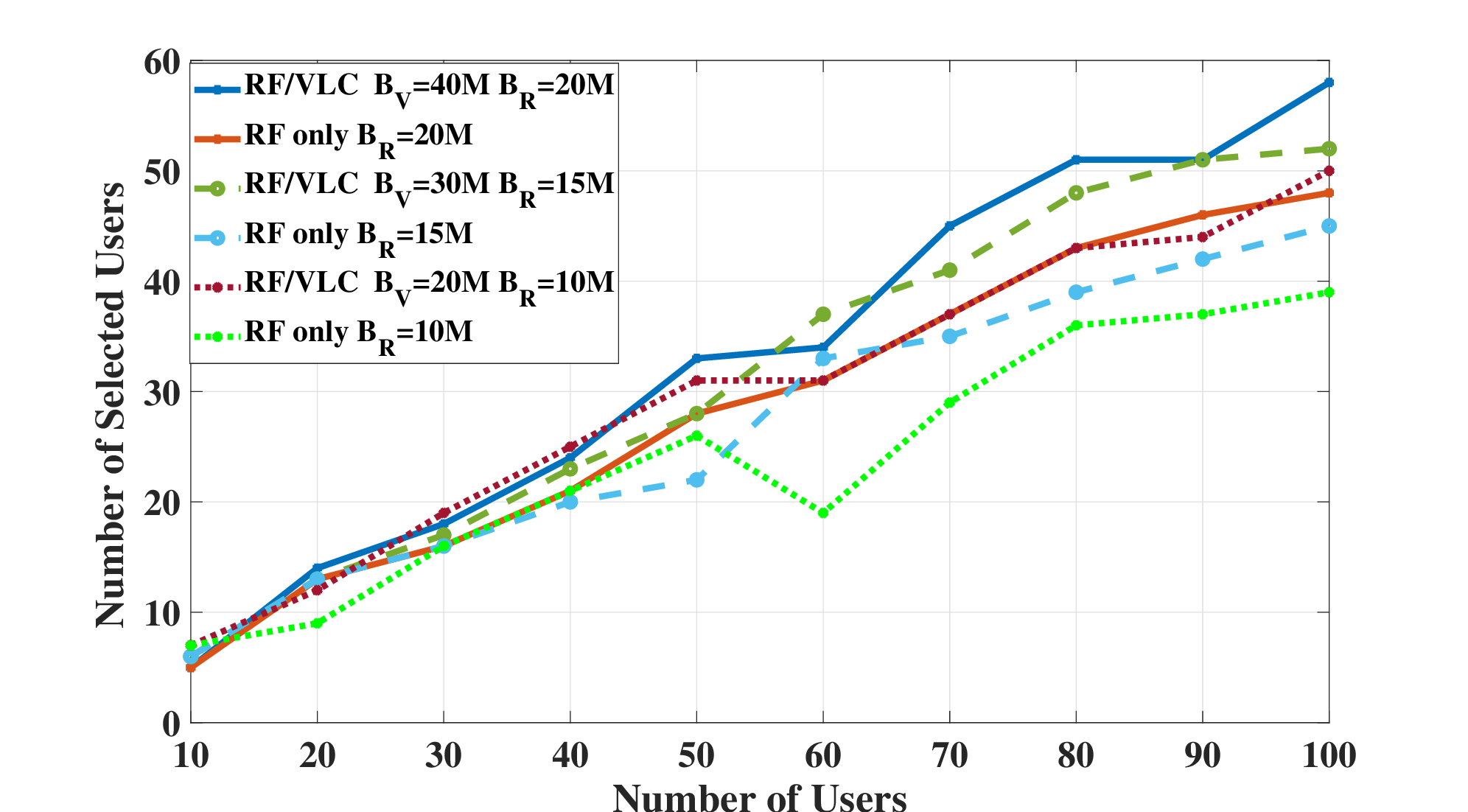}
\caption{Comparison of user selection under different bandwidth settings with different numbers of users.}
\label{fig:user}
\vspace{-0.2cm}
\end{figure}

To evaluate the performance of the proposed USBA algorithm
with different users, Fig. \ref{fig:user} shows how the number of selected
users changes as the total number of users varies. Fig. \ref{fig:user} shows that, compared to RF-only system, more users can participate
FL in the hybrid VLC/RF system. This trend is more obvious
with the increase of the number of users. In particular, when
the number of users is 50, the number of users selected by
the proposed USBA algorithm is 20\% higher than that of the benchmark. When the number of users is 100, the number of
users selected by the proposed USBA algorithm is 25\% higher
than that of the benchmark. Fig. \ref{fig:user} also compares the user
selection under different VLC and RF bandwidths. It can be
observed that the proposed USBA algorithm is better than the
benchmark under different bandwidth settings.

\vspace{-0.cm}

\begin{figure}[t]
\setlength{\abovecaptionskip}{-0.2cm}
\centering
\includegraphics[width=0.52\textwidth]{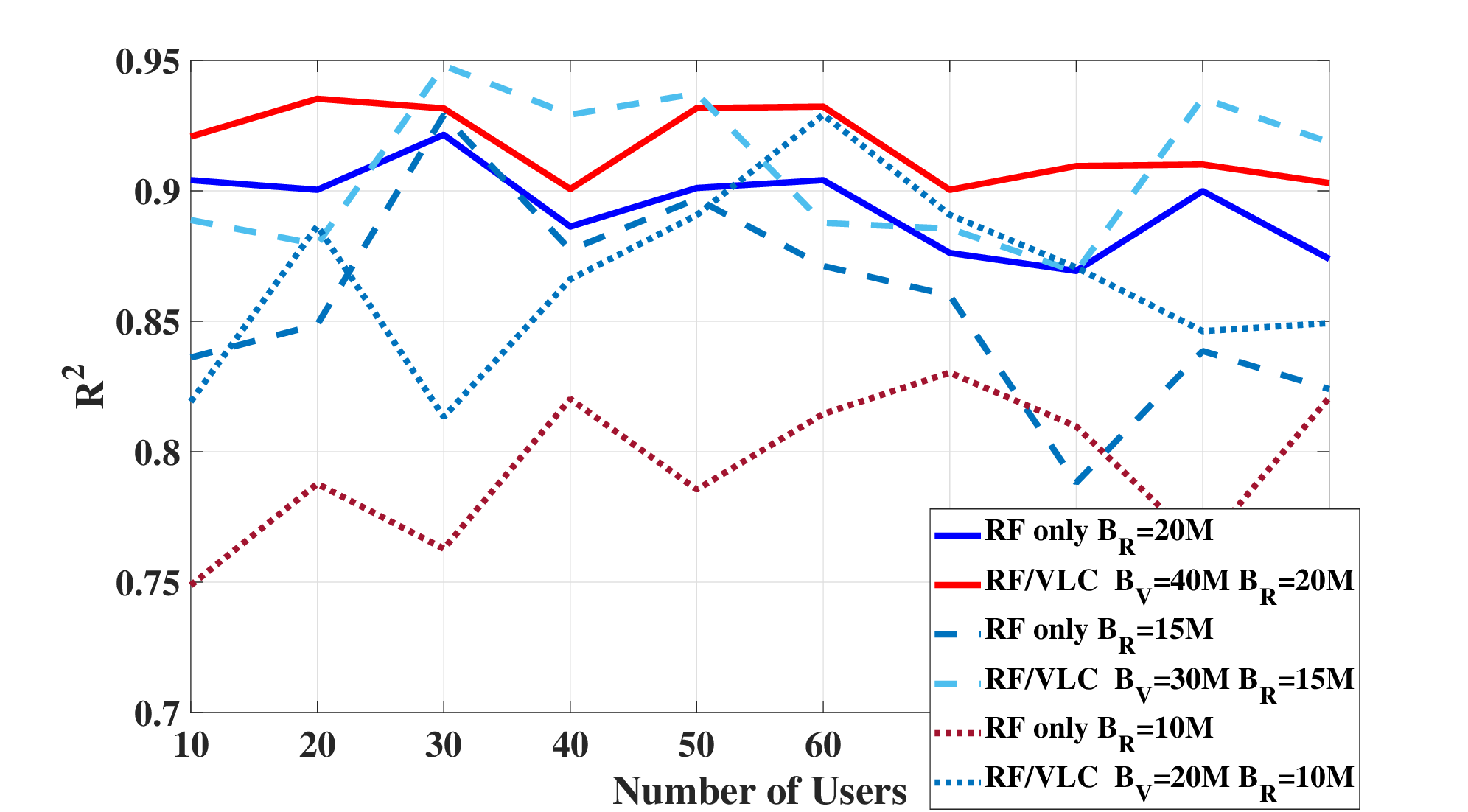}
\caption{Comparison of determination coefficients for different bandwidth settings and different numbers of users.}
\label{fig:R}
\vspace{-0.6cm}
\end{figure}

Figure \ref{fig:R} compares the values of $R^2$ of the hybrid VLC/RF
system with the RF only system under different configurations.
It can be observed that the hybrid VLC/RF system can achieve
higher $R^2$ values, which means that the proposed USBA algorithm
can make FL performance better. We can also observe
that the proposed USBA algorithm can always achieve better
performance regardless the variations of VLC/RF bandwidth.
Furthermore, the performance gain of the proposed USBA
algorithm increases with the increase of the number of total
users.

\vspace{-0.3cm}


\vspace{-0.cm}
\section{Conclusion}
This paper has proposed the introduction of VLC into conventional RF
systems for better FL performance. In particular, we have formulated
a joint user selection and bandwidth allocation problem
for FL over hybrid VLC/RF system under time, energy
and bandwidth constraints to minimize the FL training loss. We have first separated the problem into two
subproblems. The first subproblem is a user selection problem
with a given bandwidth allocation, which is solved by a traversal
algorithm. Based on the obtained user subset, the second
subproblem finds the optimal bandwidth allocation, which is
solved by a numerical method. The two subproblems are then
updated iteratively until a convergent solution is obtained.
Simulation results show that the proposed joint algorithm can
improve the number of selected users and $R^2$ by up to 20\%
and 10\%, respectively, compared with RF-only system, which
indicates that the proposed algorithm is promising for FL
training in future hybrid VLC/RF networks.

\bibliographystyle{IEEEbib}
\nocite{*}\bibliography{stimreference}

\end{document}